\documentclass[epj]{svjour}
\usepackage{graphicx}
\begin{document}

\newcommand{\eq}[1]{(\ref{#1})}

\title{The Uehling correction in muonic atoms exactly in $ Z\alpha $}
\author{E.~Yu.~Korzinin\inst{1}\thanks{E-mail: korzinin@vniim.ru}
\and V.~G.~Ivanov\inst{1,2}\thanks{E-mail: ivanov.vg@gao.spb.ru}
\and S.~G.~Karshenboim\inst{1,3}\thanks{E-mail:
savely.karshenboim@mpq.mpg.de}} \institute{ D.~I.~Mendeleev
Institute for Metrology (VNIIM), St. Petersburg 190005, Russia \and
Pulkovo Observatory, 196140 St. Petersburg, Russia\and
Max-Planck-Institut f\"ur Quantenoptik, 85748 Garching, Germany}
\date{Received: date / Revised version: date}
%
\abstract{The Uehling correction to
the energy levels is presented in terms of the hypergeometric
functions $_2F_1$. This presentation allows to derived various
asymptotics and approximations. Further applications of this method
to other atomic characteristics are also considered. \PACS{
      {36.10.Gv}{Mesonic atoms and molecules, hyperonic atoms and molecules}
      \and
      {31.30.Jv}{Relativistic and quantum electrodynamic effects in atoms and molecules}
     } 
} 
\maketitle

\section{Introduction}

Vacuum polarization effects play an important role in quantum
electrodynamics of bound states, especially in the case of
muonic and exotic atoms. They are responsible for the dominant QED
correction to the Coulomb energy levels. Recent and planned activities
on muonic hydrogen \cite{psi,hfs} demand a precision theory, which
in particular involves higher-order vacuum polarization
contributions. For instance, for muonic hydrogen such contributions
at the second and the third order were found in \cite{pach,kino}.

In this paper we consider contributions of the one-loop electronic
vacuum polarization (the so-called Uehling potential).
We intend to find such presentations of the
Uehling correction, which could be successfully applied in
higher-order calculations.

The first-order vacuum-polarization correction to energy levels in a
hydrogen-like atom is the simplest QED correction. With a known wave
function it is usually not a problem to perform a numerical
computation for any desired state. However, there is a number of
applications when an exact or approximate result in a closed
analytic form is still preferable. Among them are the estimation of
the correction for a number of levels (as a function of their
quantum numbers), the calculation of a derivative of the energy shift
(which is one of the approaches to find the related correction to the
$g$ factor of a bound electron/muon \cite{bound_g}), estimations of
relativistic and recoil corrections etc.

The energy shift induced by the Uehling potential
\begin{equation}\label{defvu}
V_{\rm U}(r) =
 -\frac{\alpha}{\pi}\;\frac{Z\alpha}{r}
 \int_0^1
 dv\,
\rho(v)\,
 \exp\left(-\frac{2m_e r}{\sqrt{1-v^2}}\right)\;,
\end{equation}
where the dispersion density function is
\begin{equation}\label{defrho}
  \rho(v)=\frac{v^2(1-v^2/3)}{1-v^2}\;,
\end{equation}
is known in a closed form in a non-relativistic approximation (see,
e.g., \cite{Pusto}) as well as exactly in $Z\alpha$. The relativistic
results for the Uehling correction to the energy were found for the
circular states in an atom with the orbiting particle of an arbitrary
mass $m$ for spin 1/2 \cite{cjp98} and 0 \cite{cjp05} exactly in $
Z\alpha $. For an arbitrary state of a Dirac particle in the Coulomb
potential the results were obtained in \cite{ejp07}.

Here and throughout the paper the relativistic units in which
$\hbar=c=1$ are applied; $m_e$ is the electron mass, $Z$ is the
nuclear charge and $\alpha$ is the fine structure constant.

The analytic results expressed in terms of the
basic integral $K_{bc}(\kappa_n)$ are cumbersome:
\begin{eqnarray}\label{kbc_def}
  K_{bc}(\kappa_n)
  &=&
  K_{1bc}(\kappa_n) - \frac{1}{3} K_{2bc}(\kappa_n)
  \,,\nonumber\\
  K_{abc}(\kappa_n)
  &=&
  \frac12 \kappa_n^{c}\,
  B\left(a+\frac12,\,1-\frac{b}{2}+\frac{c}{2}\right)
  \nonumber\\
  &\times&
  {_3F_2}\left(\frac{c}{2}, \, \frac{c}{2}+\frac12, \, 1-\frac{b}{2}+\frac{c}{2} ;\;
  \right.
  \nonumber\\
  &&
  \phantom{99999999}
  \left.
  \frac{1}{2}, \, a+\frac32-\frac{b}{2}+\frac{c}{2} ;\; \kappa_n^2\right)
  \nonumber\\
  &-&
  \frac{c}{2}\,\kappa_n^{c+1}\,
  B\left(a+\frac12, \, \frac32-\frac{b}{2}+\frac{c}{2} \right)
  \nonumber\\
  &\times&
  {_3F_2}\left( \frac{c}{2}+1,\, \frac{c}{2}+\frac12, \, \frac32-\frac{b}{2}+\frac{c}{2};\;
  \right.
  \nonumber\\
  &&
  \phantom{99999999}
  \left.
  \frac{3}{2},\, a+2-\frac{b}{2}+\frac{c}{2};\; \kappa_n^2 \right)
  \,,
\end{eqnarray}
where $_3F_2$ is the generalized hypergeometric function,\linebreak
$B(x,y)$ is the beta-function and $
  \kappa_n =   
  { Z\alpha  m}/({n m_e}) 
$.

The generalized hypergeometric function $_3F_2$ can be expanded as a
series at low $\kappa_n$, but not at high values of $\kappa_n$.
Even for $\kappa_n\ll1$ the hypergeometric series is not always
appropriate because of the increase of coefficients for $n\gg1$. A real
parameter of expansion for high $n$ and low $\kappa_n$ is
$n\kappa_n$ rather than $\kappa_n$ (see~\cite{ejp06}).

Due to all these problems it is hard to apply \eq{kbc_def} directly
and, in fact, to find asympotics in \cite{cjp98,cjp05} we used
the integral presentation instead of the explicit expression:
\begin{equation}\label{kbc_int}
  K_{bc}(\kappa_n)
  =
  \int_{0}^{1}{dv}\frac{\rho(v)}{(1-v^2)^{b/2-1}}
  \left(\frac{{\kappa_n}\sqrt{1-v^2}}{1+{\kappa_n}\sqrt{1-v^2}}\right)^{c}
  .
\end{equation}

Here we describe another way for the calculation of $K_{bc}$, which is
free of all the mentioned problems with large $\kappa_n$ and $n$
and may be applied to the higher-order perturbation theory. In
principle, the non-relativistic results for the Uehling potential
can be expressed in terms of certain elementary functions, which are
still hard to use for asymptotics etc. The method developed here is
equally efficient for non-relativistic and relativistic
calculations.

\section{Calculation of upper and lower limits of the basic
integral $K_{bc}(\kappa_n)$ as model calculation \label{s:model}}

To explain our approach, we simplify the problem and derive at first
the upper and lower limits for $K_{bc}(\kappa_n)$ instead of its
calculation.

First we change the variable in the integral presentation
(\ref{kbc_int}), introducing
$ y = \sqrt{1-v^2}$, %
and for the basic integral arrive to an equivalent form
\begin{equation}\label{kbc}
  K_{bc}(\kappa_n)
  =
  \kappa_n^{c} \int_0^1 {dy}\,  f(y) \; y^{c-b+1} \sqrt{1-y} \, ( 1+\kappa_n y )^{-c}
  \,.
\end{equation}
We note that the weight function
\begin{equation}\label{f}
  f(y) = \frac{(2+y^2)\sqrt{1+y}}{3}
\end{equation}
monotonously increases from $f_{\rm min}=f(0)=2/3\simeq0.67$ to
$f_{\rm max}=f(1)=\sqrt{2}\simeq 1.41$. It is also fruitful to
introduce a new kind of basic integrals:
\begin{equation}\label{qbc:def}
  Q_{bc}(\kappa_n)
  =
  \kappa_n^{c} \int_0^1 {dy}\,  y^{c-b+1} (1-y)^{1/2} \, [1-(-\kappa_n y) ]^{-c}
  \,.
\end{equation}
Because of the monotonous behavior of the weight function (\ref{f})
with the rest of integrand in (\ref{kbc}) being a positive factor,
we arrive to the upper and lower limits of $K_{bc}(\kappa_n)$, by
substituting $f(y)$ in \eq{kbc} by its minimal and maximal values,
respectively:
\begin{equation}\label{kminmax}
  \frac{2}{3}\, Q_{bc}(\kappa_n) \leq K_{bc}(\kappa_n)
  \leq
  \sqrt{2}\, Q_{bc}(\kappa_n)\;.
\end{equation}
Furthermore, taking advantage of the simplification of the integral,
we arrive in (\ref{qbc:def}) to the generic integral presentation of
the hypergeometric function and find
\begin{eqnarray}\label{Qk}
Q_{bc}(\kappa_n) &=&
  \kappa_n^{c} \;  B\bigl(c-b+2,3/2\bigr)
  \nonumber\\&\times&
  {}_2F_1(c,c-b+2,c-b+7/2;\;-\kappa_n)
  \,.
\end{eqnarray}
We remind that in contrast to $_3F_2$, the function $_2F_1$ has many
well-known properties and it is much easier to deal with.

The ratio of the lower and upper limits in \eq{kminmax} is obviously
a constant ${\sqrt{2}}/{3}\approx 0.5$ for any values of $b, c$ and
$\kappa_n$. That is not a trivial issue by itself, because in the wide range
from $\kappa_n\ll1$ to $\kappa_n\gg1$ the integral
$K_{bc}(\kappa_n)$ is changing by many orders of magnitude. We can
get advantage from the knowledge of the limits, e.g., by writing an
estimation
\begin{equation}\label{simple}
  K_{bc}(\kappa_n)=\bigl(1.04\pm0.37\bigr)\, Q_{bc}(\kappa_n)\;.
\end{equation}

To take further advantages of the approach above we have three key
problems to consider:
\begin{itemize}
\item an improvement in the presentation for $Q_{bc}(\kappa_n)$,
which would be easier to apply for various $\kappa_n$;
\item a derivation of estimations similar to (\ref{kminmax})
and~(\ref{simple}) for an arbitrary state which involves a number of
$K_{bc}$ integrals with different values of $b$ and $c$;
\item an improvement in the accuracy of the estimation (\ref{simple}).
\end{itemize}
Below we proceed with all three problems and consider in addition
further applications of the approach.

\section{Improved presentations for the basic integral $Q_{bc}(\kappa_n)$}

\subsection{Low-$\kappa_n$ adjusted presentation}

The presentation (\ref{Qk}) for $Q_{bc}(\kappa_n)$ allows to derive
easily a series at low $\kappa_n$. Meanwhile, its applicability is
rather limited since the coefficients $b$ and $c$ may be quite large
in the case of $n\gg1$ (see \cite{ejp06} for detail). As a first
issue, we improve a presentation for $Q_{bc}(\kappa_n)$ at low and
medium $\kappa_n$. We achieve that by applying a proper
transformation of the hypergeometric function, many of which are
well known for $_2F_1$ (see, e.g., Appendix in \cite{III}), but not
for $_3F_2$. In particular, applying Eq.~(e,5) from \cite{III} we
find
\begin{eqnarray}\label{Qk1k}
  Q_{bc}(\kappa_n)
  &=&\left( \frac{\kappa_n}{1+\kappa_n} \right)^{c} \;
  B\bigl(c-b+2,3/2\bigr)
  \nonumber\\&\times&
  {}_2F_1\left( c,3/2,c-b+7/2;\;\frac{\kappa_n}{1+\kappa_n} \right)
  \,.
\end{eqnarray}
Apparently, the advantage of this presentation is that the argument
of the hypergeometric function is always below unity (since
$\kappa_n$ is positive). We note the appearance of pre-factor
$({\kappa_n}/({1+\kappa_n}))^{c}$ in (\ref{Qk1k}), which carries
most of the changes in the value of $K_{bc}(\kappa_n)$, while the
remaining factor is a slow-changing smooth function of $\kappa_n$
(see Sect. 6.3 in \cite{ejp06}; cf. Eq.~(\ref{Q11k}) in this paper).

For an arbitrary state we discuss the problem in
Sect.~\ref{s:states}, and here we present examples for the
non-relativistic case with the circular states ($l=n-1$) only, for
which the whole correction to the energy is determined by a single
$K_{bc}$ integral \cite{cjp98,ejp06} with $b=2$ and $c=2n$. In this
particular case, we note another advantage of presentation
(\ref{Qk1k}). The convergency of the hypergeometric series
\[
_2F_1(\alpha,\alpha^\prime,\beta;z)
=\frac{\Gamma(\beta)}{\Gamma(\alpha)\Gamma(\alpha^\prime)}\,
\sum_{k=0}^\infty{\frac{\Gamma(\alpha+k)\Gamma(\alpha^\prime+k)}{\Gamma(\beta+k)}
\,\frac{z^k}{k!}}
\]
is different in terms of $\kappa_n$ for different presentations.
While the series in \eq{Qk} is finite only for a limited
range\footnote{E.g., for $c\gg1$ and $b=2$ the series is convergent
for $c\,\kappa_n/2<1$ \cite{ejp06}.} of low $\kappa_n$, the series in
the presentations (\ref{Qk1k}) is convergent for
$0<\kappa_n<\infty$.

Indeed, the speed of the convergency of the series in \eq{Qk} and
\eq{Qk1k} is also quite different and in particular at low
$\kappa_n$ the convergency is much faster for high $n$ for $_2F_1$
in \eq{Qk1k} than in \eq{Qk}. That is because in the former case for
high $n$ the parameters are: $\alpha\sim\beta\sim 2n$ and
$\alpha^\prime\sim 3/2$, while in the latter presentation:
$\alpha\sim\alpha^\prime\sim\beta\sim 2n$.

\begin{figure}
\begin{center}
\includegraphics[width=0.4\textwidth]{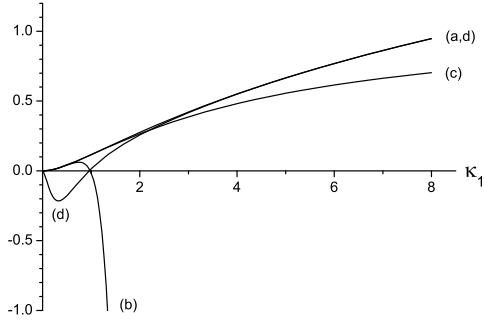}
\end{center}
\caption{Comparison of the convergence of the hypergeometric
function in relations~(\ref{Qk}), (\ref{Qk1k}) and \eq{Q11k} for the
$1s$ state: (a) -- the exact function $Q_{bc}(\kappa_1)$, (b) -- the
first five terms of expansion of $_2F_1$ in \eq{Qk}, (c) -- the same
for \eq{Qk1k}, (d) -- the same for \eq{Q11k}.}
\label{fig:kbc_2f1_n1}
\end{figure}

\begin{figure}
\begin{center}
\includegraphics[width=0.4\textwidth]{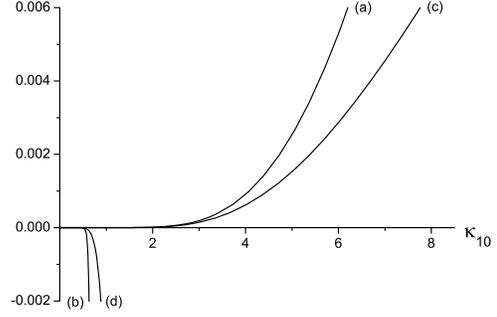}
\includegraphics[width=0.4\textwidth]{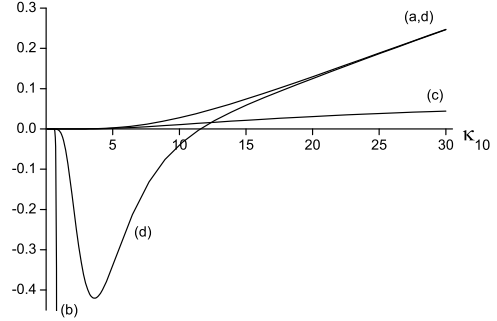}
\end{center}
\caption{Comparison of the convergence of the hypergeometric
function in relations (\ref{Qk}), (\ref{Qk1k}) and \eq{Q11k} for the
circular state with $n=10$: (a) -- the exact function
$Q_{bc}(\kappa_{10})$, (b) -- the first five terms of expansion of
$_2F_1$ in \eq{Qk}, (c) -- the same for \eq{Qk1k}, (d) -- the same
for \eq{Q11k}.} \label{fig:kbc_2f1_n10}
\end{figure}

The results for $Q_{bc}(\kappa_n)$ related to the ground state and
to a circular state at $n=10$ are presented in
Figs.~\ref{fig:kbc_2f1_n1} and~\ref{fig:kbc_2f1_n10}, respectively.
To check how fast the convergency is we consider only the first five
terms of the hypergeometric series for each presentation of $_2F_1$.

The behavior of the five-term approximation in (\ref{Qk}) is very
notable (cf. also Fig.~\ref{fig:frelkzi}). As explained in
\cite{ejp06,cjp07}, the actual parameter of the naive low-$\kappa_n$
expansion, based on (\ref{Qk}), is $n\kappa_n$ and at
$n\kappa_n\to1$ the error of the related five-term approximation
becomes very high. In contrast to that, the presentation (\ref{Qk1k})
allows an improved low-$\kappa_n$ approximation which is successful
at least up to $\kappa_n\sim 1$.

\begin{figure}
\begin{center}
{\includegraphics[width=0.35\textwidth]{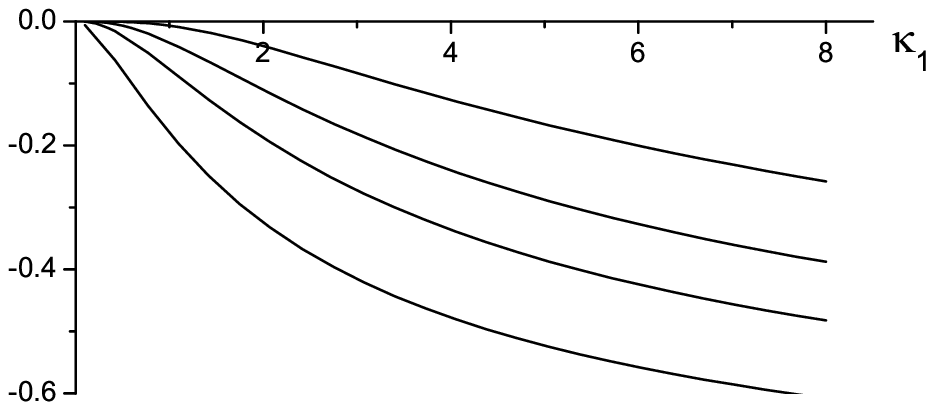}}
\includegraphics[width=0.35\textwidth]{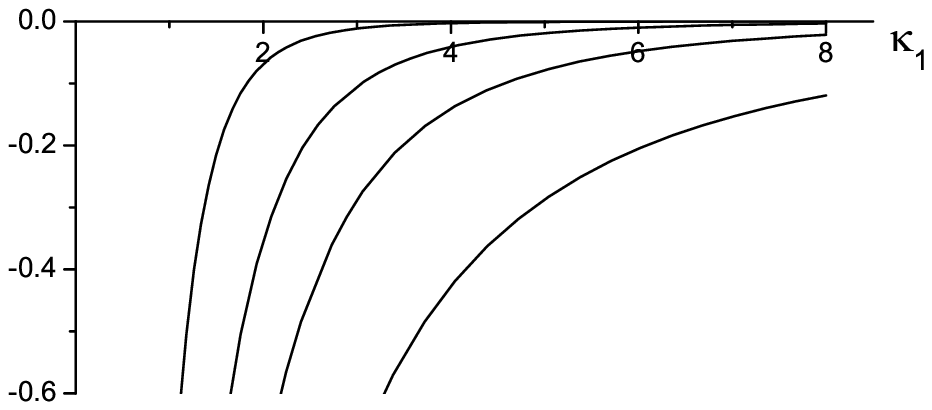}
\end{center}
\caption{Relative errors for the hypergeometric function expansions
in relations (\ref{Qk1k}) (top) and (\ref{Q11k}) (bottom) as
compared with the exact value for different number of non-trivial
terms in the expansion (from bottom to up: 1,2,3,5 terms). The
results are for the $1s$ state.} \label{fig:kbc_2f1_n1_terms}
\end{figure}

We can also check how many terms of the hypergeometric series we
need to reach a proper level of accuracy. The results are collected
in Figs.~\ref{fig:kbc_2f1_n1_terms} and~\ref{fig:kbc_2f1_n10_terms}.
One can note that the 5-term approximations are quite successful in
a broad area. E.g., as seen in
Figs.~\ref{fig:kbc_2f1_n1}--\ref{fig:kbc_2f1_n10_terms}, the
improved low-$\kappa_n$ expansion is accurate up to $\kappa_n\simeq
1$.

\subsection{High-$\kappa_n$ adjusted presentation\label{ss:hkappa}}

In contrast to the presentation (\ref{kbc_def}) for $K_{bc}$ with
$_3F_2$, the presentations (\ref{Qk1k}) for $Q_{bc}$ with $_2F_1$
are easy to adjust for the high-$\kappa_n$ case.

To derive a presentation efficient at high $\kappa_n$, we apply to
\eq{Qk1k} the transformation (e,7) from \cite{III} and find
\begin{eqnarray}\label{Qbc11k}
  Q_{bc}(\kappa_n)
  &=&
 \left( \frac{\kappa_n}{1+\kappa_n} \right)^c
  \times
  \Biggl[
    B\bigl(2-b,3/2\bigr)
    \nonumber\\&&\times\;
    {}_2F_1\left( c,3/2,b-1;\;\frac{1}{1+\kappa_n} \right)
  \\&&+
    (1+\kappa_n)^{b-2}
    B\bigl(-2+b,c-b+2\bigr)
  \nonumber\\&&\times\;
    {}_2F_1\left( c-b+2,7/2-b,3-b;\;\frac{1}{1+\kappa_n} \right)
  \Biggr]\,.
  \nonumber
\end{eqnarray}

We note that for the case of integer $b$, which actually takes place for any
state, the expression contains a singularity and should be properly
regularized. In a general case we apply the substitute $b\to
b+\epsilon$ in \eq{Qk1k}, which is obviously valid, next apply the
transformation (e,7) from \cite{III} and then consider a limit of
$\epsilon\to0$ in (\ref{Qbc11k}). That delivers us a finite
well-determined result. Eventually, the result for $b=2$ reads
\begin{eqnarray}\label{Q11k}
  Q_{2c}(\kappa_n)
  &=&
  \left( \frac{\kappa_n}{1+\kappa_n} \right)^c
  \nonumber\\&&\times\;
  \biggl\{\frac{\partial}{\partial\epsilon}\,
  {}_2F_1\left( c-\epsilon,\frac32-\epsilon,
  1-2\epsilon,\frac{1}{1+\kappa_n}
  \right)\bigg\vert_{\;\epsilon=0}
\nonumber
  \\&&+\;
  \Bigl[ \ln(1+\kappa_n) -\psi(3/2)-\psi(c)+2\psi(1) \Bigr]
 \nonumber\\
  &&\times{}_2F_1\left(c,\frac32,1,\frac{1}{1+\kappa_n} \right)
   \biggr\}\;,
\end{eqnarray}
where $\psi(x)$ is the logarithmic derivative of Euler's gamma
function. A comparison of approximation of this presentation by a
partial sum of the hypergeometric series is summarized in
Figs.~\ref{fig:kbc_2f1_n1} and \ref{fig:kbc_2f1_n10}. The results
for general value of $b$ is presented in App.~\ref{as:hkappa}.

\begin{figure}
\begin{center}
\includegraphics[width=0.35\textwidth]{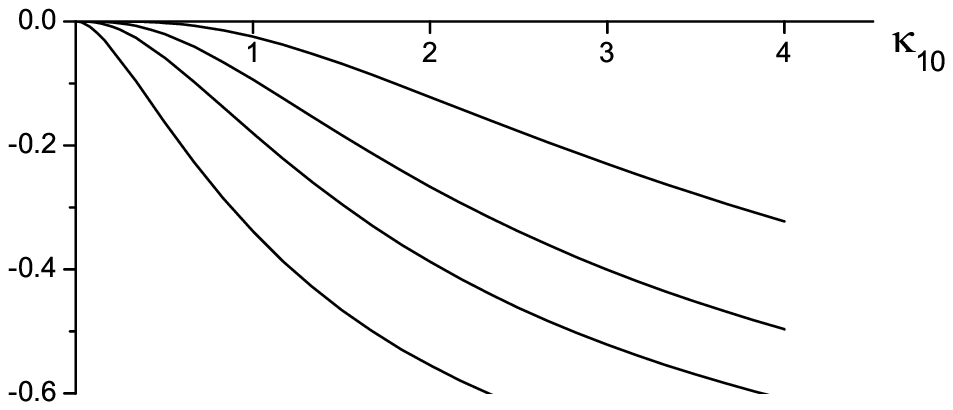}
\includegraphics[width=0.35\textwidth]{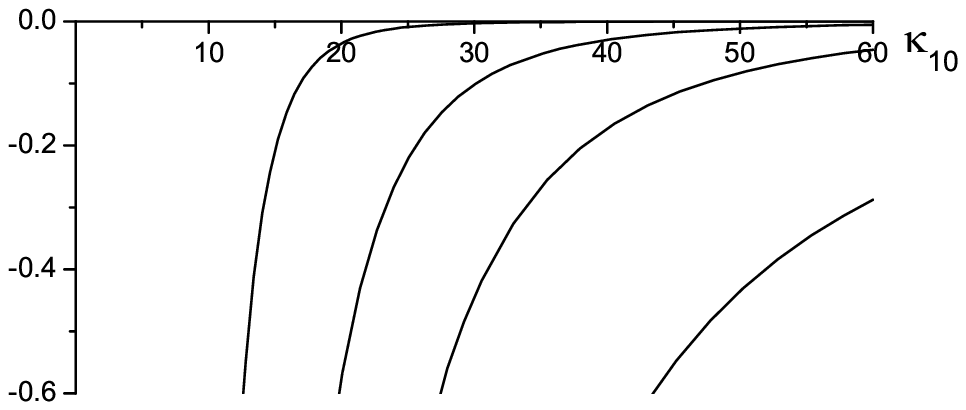}
\end{center}
\caption{Relative errors for the hypergeometric function expansions
in relations (\ref{Qk1k}) (top) and~(\ref{Q11k}) (bottom) as
compared with the exact value for different number of non-trivial
terms in the expansion (from bottom to up: 1,2,3,5 terms). The
results are for the circular state with $n=10$.}
\label{fig:kbc_2f1_n10_terms}
\end{figure}

Obviously the high-$\kappa_n$ asymptotics (cf.~\cite{cjp98}) should
contain $\ln\kappa_n$, which follows from well-known asymptotics of
the vacuum polarization. The derived presentation allows to separate
the logarithmic contribution explicitly.

With appearing of the $_2F_1$ function of the argument
$1/(1+\kappa_n)$ one can consider the approximation of $_2F_1$ by a
few first terms of the hypergeometric series (see
Figs.~\ref{fig:kbc_2f1_n1_terms} and~\ref{fig:kbc_2f1_n10_terms}).
We see that $c=2n$ and other parameters of $_2F_1$ are of the order
of unity, therefore for high $\kappa_n$ the expansion is effectively
done in $n/\kappa_n$ (cf. \cite{ejp06}).

As long as we deal with (\ref{kminmax}) and~(\ref{simple}), we can
present the result for any state with combinations of
$Q_{2c}(\kappa_n)$. However, improvements of
estimation~(\ref{simple}), which we consider below, involve various
cases of $_2F_1$ and the transformation becomes more complicated
(see App.~\ref{as:hkappa} for more detail).

\section{Estimations for corrections to the energy shifts \label{s:states}}

The estimations above are achieved for the basic integral
$K_{bc}(\kappa_n)$, while the results for any state except
circular ones contain a set of such integrals. In particular, the
non-relativistic result for an arbitrary $nl$ state is of the form
\cite{ejp06}
\begin{equation}
\Delta E_{nl} = \frac{\alpha}{\pi}\frac{( Z\alpha )^2m}{n^2}
F_{nl}(\kappa_n)
 \;,
\end{equation}
\begin{eqnarray}\label{ENRnl}
    F_{nl}(\kappa_n) =
  &-&
  \frac{(n+l)!}{n_r!}
  \sum_{i=0}^{n_r}
  \frac{1}{i!(2l+i+1)!}
  \\
  &\times&
  \left( \frac{n_r!}{(n_r-i)!} \right)^2
  \frac{K_{2{n_r},2n}(\kappa_n)}{\kappa_n^{2(n_r-i)}}
  \nonumber\,,
\end{eqnarray}
where $n_r=n-l-1$.

This expression is a sum of terms with the same sign and we can
apply to each $K_{bc}$ separately the approach based on our
approximation for $f(y)$. Before doing that, we like to mention that
the approach can be also applied to various other representation, such
as a sign-alternating series \cite{ejp07}
\begin{eqnarray}\label{ERnljLim}
  F_{nl}(\kappa_n)
  &=&
  - \sum_{i=0}^{n_r}\sum_{k=0}^{n_r} \frac{(-1)^{i+k}n_r!(n+l)!}{i!(n_r-i)!k!(n_r-k)!}
  \\&&\times\;
   \frac{(2l+i+k+1)!}{(2l+i+1)!(2l+k+1)!}
    K_{2,2l+i+k+2}(\kappa_n)
  \nonumber
\end{eqnarray}
or a presentation with derivatives of $K_{bc}$ \cite{ejp06}
\begin{eqnarray}\label{onemore}
  F_{nl}(\kappa_n)&=&
  \frac{(n+l)!}{n_r!(2n-1)!}
  \sum_{i=0}^{n-l-1}
  \frac{1}{(2l+i+1)!}
  \,
  \frac{1}{i!}
\nonumber\\
&\times&\left(\frac{1}{\kappa_n}\right)^{2(n_r-i)}
 \left( \frac{n_r!}{(n_r-i)!} \right)^2
 \left( \kappa_n^2 \frac{\partial}{\partial\kappa_n} \right)^{2(n_r-i)}
 \nonumber\\&&\times
  \,
  \kappa_n^{2(l+i+1)}
  \,
  \left( \frac{\partial}{\partial\kappa_n} \right)^{2(l+i)}
  \frac{F_{10}(\kappa_n)}{\kappa_n^2}
  \label{fnlf10}
  \,,
\end{eqnarray}
which expresses the correction for an arbitrary non-re\-la\-ti\-vis\-tic
state in terms of the result for the $1s$ state, which is well known
(see two previous equations at $n=1$, $l=0$) \cite{Pusto}. One more
presentation for $F_{nl}(\kappa_n)$ can be found in \cite{soto}.

Validity of estimations presented in Sect.~\ref{s:model} and
similar can be easily proved not through any explicit analytic
presentations of $F_{nl}$, but via related original expressions for
the energy shift before any integrations are taken.

Such an expression in both relativistic (Dirac's) and
non-relativistic case is of the form
\begin{eqnarray}\label{general}
\Delta E &=&
 -\frac{\alpha(Z\alpha)}{\pi}
 \int
 d^3 r
 \int_0^1
 dv \,
 \rho(v)
 \nonumber\\&\times&
 P(r) \,
 \frac{1}{r}
 \exp\left( -\frac{2 Z\alpha  m_e r}{\sqrt{1-v^2} } \right)
 \;.
\end{eqnarray}
Here $v$ is the spectral parameter for the Schwinger presentation
\cite{Schwinger} of the vacuum polarization, $\rho(v)$ is the
spectral function, which in the one-loop case is defined in
(\ref{defrho}), while $P(r)$ is a positive weight function which
describes the density of the distribution of the electric charge of the
atomic bound particle. The latter in the non-relativistic case reads
\[
  P_{\rm NR}(r) =  \vert\Psi(r)\vert^2
  \,,
\]
where $\Psi(r)$ is the Schr\"odinger wave function of the related
atomic state, while in the relativistic case it is defined as
\[
  P_{\rm Rel}(r) =  \vert f(r)\vert^2 + \vert g(r)\vert^2
  \,,
\]
where $f(r)$ and $g(r)$ are the radial parts of the upper and lower
components of the Dirac wave function.

All evaluations in Sect.~\ref{s:model} are based on a certain
manipulation with $\rho(v)$ and on the fact that the $v$-integrand
is positive. The same can be seen in \eq{general}. That means that for
any state we can replace $f(y)$ (which is a result of a certain
transformation of $\rho(v)$) by its minimum and maximum and in this
way we arrive at the upper and lower limits for the whole $F_{nl}$.
To derive such limits we may apply any presentation for $F_{nl}$
listed above, by substitute $K_{bc}$ in the right-hand side for its
upper and lower limit.

That would not be clear from the point of view of the presentations
above by themselves, which include positive and negative
contributions as well as derivatives. However, substituting $f(y)$
by $f(0)$ (or $f(1)$), we should arrive at such a limitation. That
is obvious, as seen from the consideration above for any
presentation without derivatives (such as \eq{ENRnl} and
\eq{ERnljLim}). Concerning (\ref{onemore}), we remind that it is
deduced from a presentation without derivatives by using certain
recurrent relations \cite{ejp06} (see also \cite{cjp98}). Meanwhile,
the relations are maintained by the shape of (\ref{kbc_int}) for any
spectral function $\rho(v)$ and thus allow any substitution for
$f(y)$.

In the presentation (\ref{ERnljLim}) $b=2$ for all
integrals and $c$ can take various values, while in the
case of (\ref{ENRnl}) the situation is opposite: $c=2n$ and $b$
varies. The consideration above and in particular the
high-$\kappa_n$ expansion in Sect.~\ref{ss:hkappa} is derived only for
$b=2$. Applying this approach to sum (\ref{ERnljLim}), one can arrive to more
complicated identities (see App.~\ref{as:hkappa} for
more detail).

\section{Presentation of the weight function $f(y)$ as infinite
series exactly and its direct approximation by polynomials}

Above we found that simplifying the function $f(y)$ in (\ref{kbc})
(in particular, replacing it by a constant) we succeeded to simplify
the integral and present the result in substantially simpler terms
than in (\ref{kbc_def}).

\begin{figure}
\begin{center}
\includegraphics[width=0.35\textwidth]{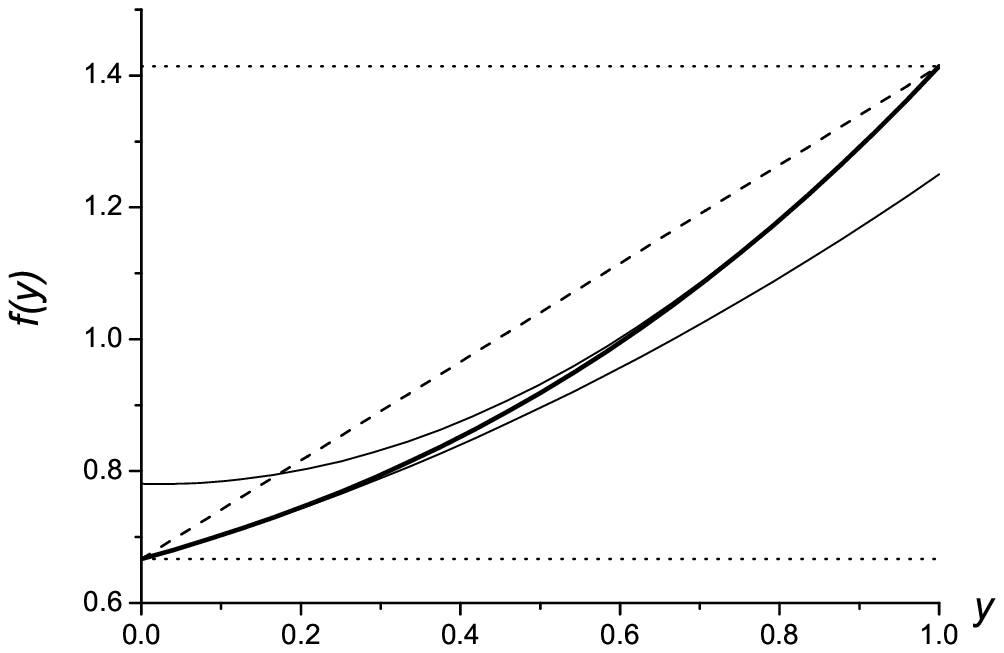}
\includegraphics[width=0.35\textwidth]{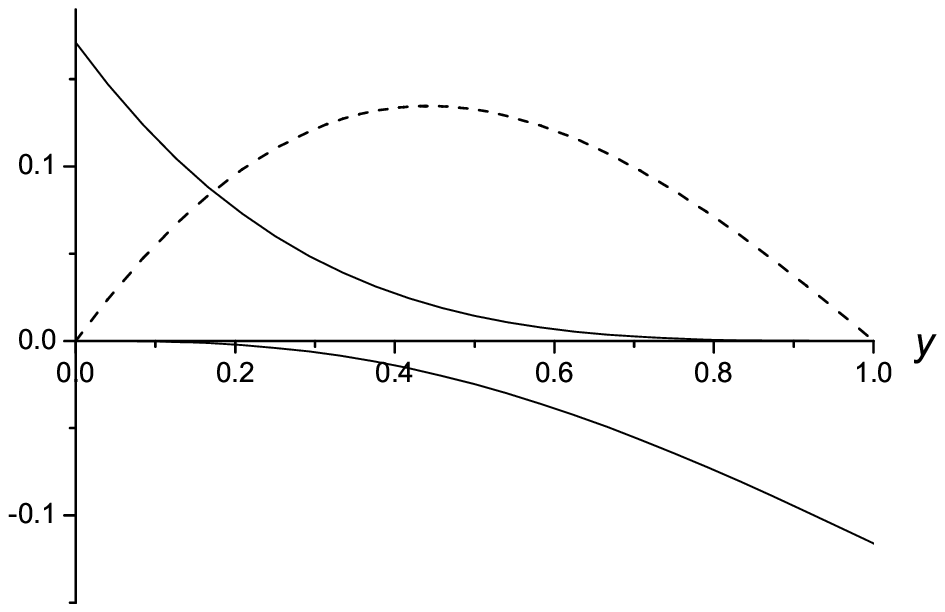}
\end{center}
\caption{Top: plots of $f(y)$ \eq{kbc} (bold line), its partial
sums with first the first three terms for series (\ref{eithery}) and
(\ref{either1y}) (thin lines) and the linear approximation~(\ref{flin})
(dashes). The maximal and minimal values of $f(y)$ are denoted by
the dotted lines. Bottom: the relative errors of the three-term
partial sum (thin lines) and the linear approximation (dashes).}
\label{fig:fy3t}
\end{figure}

In this section we consider a presentation of $f(y)$ by infinite
series in terms of either form
\begin{eqnarray}
\label{eithery}
 f(y)  &=&
  \sum_{k=0}^{\infty} c_k \,y^{k}
\,,\\
\label{either1y}
f(y) &=&\sum_{k=0}^{\infty} c^\prime_k\,
(1-y)^{k}
   \;.
\end{eqnarray}
The finite series of this kind can already be used as an
approximation (see, e.g., Fig.~\ref{fig:fy3t}), but direct
approximations lead to even simpler and more accurate presentations
for the energy shift (see Sect.~\ref{ss:poly} below).

Here we consider series and approximations for the basic integral
$K_{bc}(\kappa_n)$, leaving discussions of uncertainties of partial
sums of series for App.~\ref{as:unc}.

Before considering any particular expansion or approximation based
on (\ref{eithery}) and (\ref{either1y}), we note that the weight
function $f(y)$ is positive as well as the integrand, as discussed in
Sect.~\ref{s:states}.

In particular, we start from
\[
 \Delta E = - \int_0^1{dy}f(y)R(y)
 \,,
\]
where $R(y)$ is a certain positive function. Considering
the approximation of $f(y)$ by $\widetilde{f}(y)$ and introducing the correction
factor
\[
\widetilde{f}(y)=N(y)\,f(y),
\]
we relate the approximation of the energy shift
\[
\widetilde{\Delta E} = - \int_0^1{dy}\widetilde{f}(y)R(y)
\]
to the exact value as
\begin{equation}\label{master_est}
\widetilde{\Delta E} = -
N(\overline{y})\,\int_0^1{dy}f(y)R(y)=N(\overline{y})\,\Delta E\;,
\end{equation}
where $\overline{y}$ is a certain unknown intermediate value
$0<\overline{y}<1$. To make a successful approximation it is enough
to approximate the function $f(y)$ in such a way that it does not
exceed a certain margin: $\vert N(y)-1\vert\leq \delta$. So, the
relative uncertainty of the approximation is at least below
$\delta$. That is valid for arbitrary $R(y)$ and in particular for
the energy shift for an arbitrary state in a muonic atom.

\subsection{Presentation of $f(y)$ by an infinite series and the results for the integral $K_{bc}(\kappa_n)$}

Presenting $f(y)$ as a Taylor series, one can find the coefficients of
either series~(\ref{eithery}) and~(\ref{either1y}), which are
summarized in Table~\ref{tab:cck}.

\begin{table}
 \begin{center}
 \begin{tabular}{c||l|l}
 \hline $k$ &   $c_k$ &  $c^\prime_k$ \\
 \hline
 0 &2/3 &$\sqrt{2}$ \\
 1 &1/3 &$-\frac{11}{6\sqrt{2}}$ \\
 2 &1/4 &$\frac{15}{16\sqrt{2}}$ \\
 3 &5/24 &$-\frac{9}{64\sqrt{2}} $\\
 4 &-13/192 &$-\frac{47}{3072\sqrt{2}}$ \\
 5 &5/128 &$-\frac{15}{4096\sqrt{2}}$ \\
 $k\rightarrow\infty$ &$(-1)^{k+1}e^{7/2}\frac{1}{2\sqrt{\pi}}\,\frac{1}{k^2}$ &
 $-e^{7/2}\frac{1}{\sqrt{2\pi}}\,\frac{1}{2^k k^2}$
\\[1ex]
\hline
&&\\[-1ex]
$k$
& $\frac{(-1)^{k+1}\bigl(2k^2-6k+5\bigr)}{4\sqrt\pi}$
& $-\frac{\bigl(4k^2-8k+15\bigr)}{2^{k+2}\sqrt{2\pi}}$
\\[1ex]
& $\phantom{9999999}\times\frac{\Gamma(k-5/2)}{k!}$
& $\phantom{9999}\times\frac{\Gamma(k-5/2)}{k!}$
\\[1ex] \hline
\end{tabular}
\end{center}
\caption{Coefficients of the Taylor series for $f(y)$ expansions
(\ref{eithery}) and~(\ref{either1y}).} \label{tab:cck}
\end{table}

Applying the series (\ref{eithery}) to the integral (\ref{kbc})
leads to an infinite sum
\begin{equation}\label{Kbcser0}
  K_{bc}(\kappa_n)
  =\sum_{k=0}^{\infty}
\,c_k\,Q_{b-k,c}(\kappa_n)
  \;,
\end{equation}
where one can use for $Q_{b-k,c}$ an appropriate presentation (see,
e.g.,  (\ref{Qk}), (\ref{Qk1k}), (\ref{Q11k}) etc.).

A similar evaluation of (\ref{either1y}) leads to the result
\begin{eqnarray}\label{Kbcser1}
  K_{bc}(\kappa_n)
  &=&
  \left( \frac{\kappa_n}{1+\kappa_n} \right)^{c}
  \,
  \sum_{k=0}^{\infty}
   c^\prime_k\,
   B\left(c-b+2,k+\frac32\right)
\nonumber\\&\times& {}_2F_1\left(
c,\frac32+k,c-b+\frac72+k;\,\frac{\kappa_n}{1+\kappa_n} \right)
  .
\end{eqnarray}
A presentation similar to (\ref{Qk}) and (\ref{Q11k}) can be also
derived.

An important property of the both series for $f(y)$ is that the
coefficients are quite regular. Their absolute values decrease with
$k$. While the coefficients in \eq{eithery} are regularly
sign-alternating (except of few first terms), coefficients in
\eq{either1y} are all negative (except of a few first terms). Such a
regular structure allows a simple conservative estimation of
accuracy of a partial sum of either series (see App.~\ref{as:unc}).
To conservatively estimate the remainder of the series, we do the
estimation prior the $y$-integration, dealing only with a certain
expansion of function $f(y)$, which is indeed universal for any
state.

\subsection{Polynomial approximations\label{ss:poly}}

The series above allow in principle to reach any accuracy. However,
one very seldom needs the accuracy substantially better than 0.01\%,
and it may be more fruitful not to expand the function, but to
approximate it.

To obtain an approximation for $K_{bc}(\kappa_n)$ we need to
approximate successfully $f(y)$, e.g., in a form
\begin{eqnarray}\label{fy}
  f(y) &\simeq& \tilde f(y) = f(0) \cdot(1-y) + f(1)\cdot y
  \nonumber\\
  &+& y(1-y)\cdot \bigl(d_0 + d_1 y + d_2 y^2 + \dots\bigr)
\end{eqnarray}
and to tune the coefficients $d_k$ to minimize the value
\[
  \delta = \max\limits_{0\le y \le 1} \vert N(y)-1\vert\;.
\]

For instance, a very rough linear approximation
\begin{equation}\label{flin}
   \tilde f(y) = f_0(y) = f(0)\cdot(1-y) + f(1)\cdot y
\end{equation}
is already compatible with the partial three-term sums for either series
considered above (see Fig.~\ref{fig:fy3t}). The results of various
approximations are summarized in Table~\ref{tab:ap}.

\begin{table}
 \begin{center}
 \begin{tabular}{c|l|l|l|l}
 \hline  & $f_0$  &  $f_1$ & $f_2$  &  $f_3$   \\
 \hline
$d_0$& 0 & $-0.477333$ & $-0.418767$ & $-0.41490$\\
$d_1$& 0 & 0           & $-0.135733$ & $-0.15737$\\
$d_2$& 0 & 0           & 0           & $~0.02412$\\
\hline
$\delta$&$13\%$ & $0.7\%$&$0.03\%$& $0.003\%$\\
\hline
\end{tabular}
\end{center}
\caption{Coefficients of various approximations of $f(y)$ in
(\ref{fy}). The approximations are denoted as $f_i(y)$.}
\label{tab:ap}
\end{table}

Including one more coefficient we improve the accuracy substantially.
Apparently, we can continue and reach any required accuracy. We
already know that a successful polynomial approximation is possible
because of the existence of series~(\ref{eithery})
and~(\ref{either1y}), meanwhile here we use a more direct strategy
to minimize the uncertainty of an approximation with a fixed number
of terms.

\begin{figure}
\begin{center}
\includegraphics[width=0.35\textwidth]{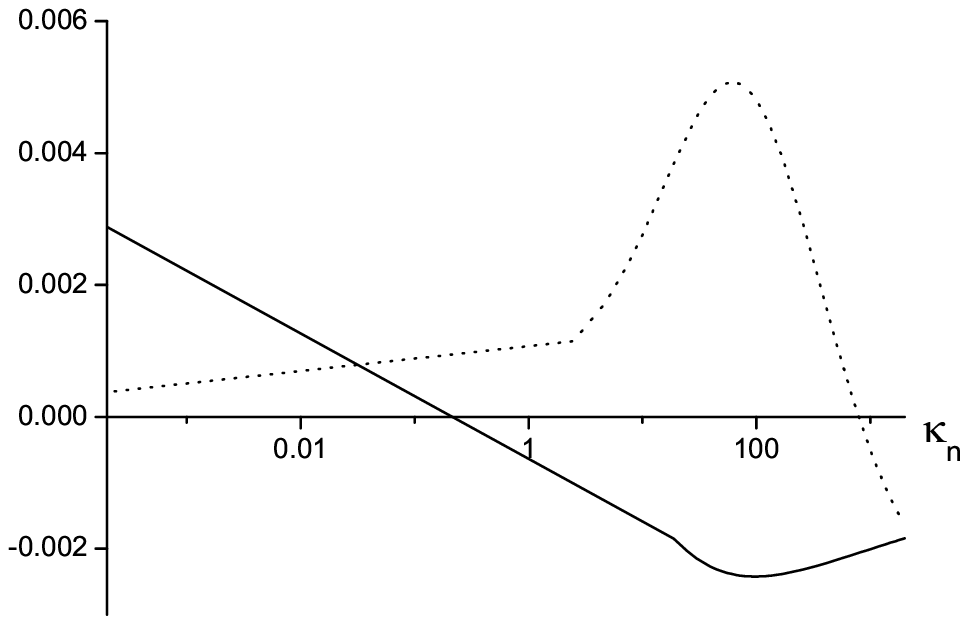}
\includegraphics[width=0.35\textwidth]{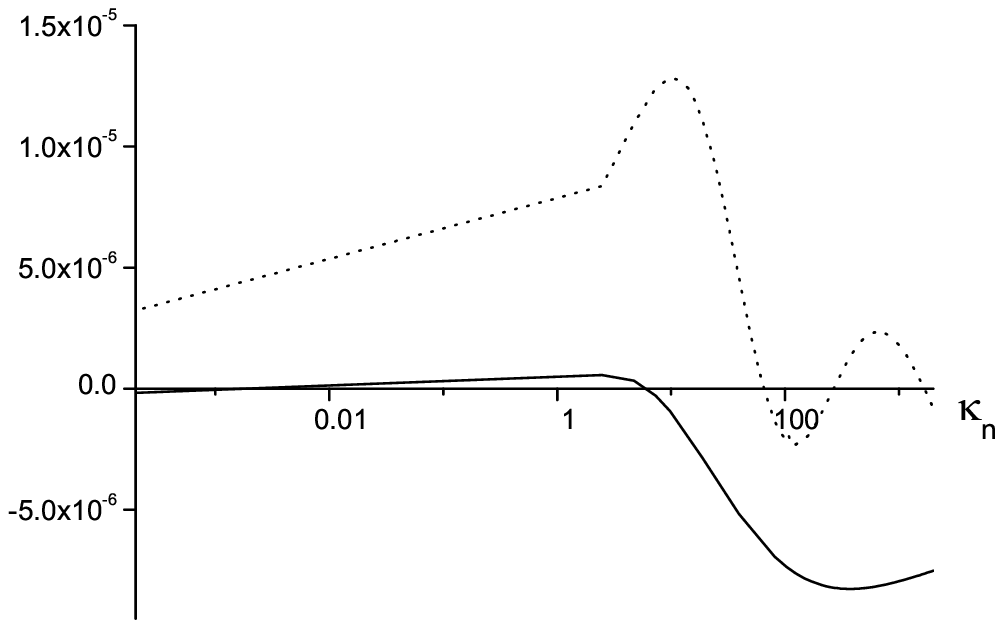}
\end{center}
\caption{Top: the relative error of the approximative functions
$K_{2,2n}(\kappa_n)$, corresponding to function $\tilde f(y)=
f_1(y)$, for $n=1$ (solid line) and $n=100$ (dots). Bottom: the same
for function $\tilde f(y)=f_3(y)$.}
\label{fig:approx}
\end{figure}

Applying to (\ref{kbc}) the approximation
\[
 \tilde{f}(y) =    \sum_{k=0}^{N} \tilde{d}_k \,y^{k}\;,
\]
where
\begin{eqnarray}
\tilde{d}_0&=&f(0),\quad\tilde{d}_1=f(1)-f(0)+d_0,\nonumber\\
\tilde{d}_k&=&{d}_{k-1}-{d}_{k-2}~~{\rm for~}k\geq2\;,\nonumber
\end{eqnarray}
we obtain the corresponding approximation for
$K_{bc}$
\begin{equation}\label{kbc_approx}
  \widetilde{K}_{bc}(\kappa_n)
  =\sum_{k=0}^{N}
\,\tilde{d}_k\,Q_{b-k,c}(\kappa_n)
  \;.
\end{equation}
We expect our estimation of accuracy to be rather conservative.
A fractional error of the approximation $\tilde f(y)$ is plotted in
Fig.~\ref{fig:approx} for the non-relativistic basic integral
$K_{2c}$, which, we remind, completely determines the energy shift
for the circular state with $n=c/2$.

\section{Further applications}

Above we have developed a method for calculation of the
non-relativistic Uehling correction to the energy. Below we discuss
possible further applications. Prior to the discussion we have to note
that the application of some results, such as a presentation by
the infinite series~(\ref{Kbcser0}) and~(\ref{Kbcser1}), is quite
straightforward. Some other need more consideration to estimate the
fractional uncertainty. That concerns, in particular, the
approximation (\ref{kbc_approx}). When the complete $y$ integrand
does not change sign, we can apply an estimation from
Table~\ref{tab:ap}, otherwise we consider our approach as a `good'
approximation, the uncertainty of which is to be revisited in each
particular case.

\subsection{Relativistic Uehling correction for a Dirac particle}

The relativistic expression for the energy shift of the $nl_j$ state
in a hydrogen-like atom with a Dirac particle (the electron, muon or
antiproton) reads as \cite{ejp07}
\begin{equation} \Delta E_{nlj} = \frac{\alpha}{\pi}\frac{( Z\alpha
)^2}{n^2} F_{nlj}(\widetilde{\kappa}_n)
 \;,
\end{equation}
\begin{eqnarray}\label{ERnlj}
  F_{nlj}(\widetilde{\kappa}_n)
  &=&
  -\frac{n^2\eta^2}{( Z\alpha )^2}
  \,
  \frac{\Gamma(2\zeta+n_r^\prime+1) n_r^\prime!}{\frac{ Z\alpha }{\eta}-\nu}
  \nonumber\\&\times&
  \sum_{i,k=0}^{n_r^\prime}
  \frac{(-1)^{i+k}}{i!(n_r^\prime-i)!k!(n_r^\prime-k)!}
  \nonumber\\&\times&
  \frac{\Gamma(2\zeta+i+k)}{\Gamma(2\zeta+i+1) \Gamma(2\zeta+k+1)}
  \nonumber\\&\times&
  \Biggl\{
  m\left[
    \left( \frac{ Z\alpha }{\eta}-\nu \right)^2
    + (n_r^\prime-i)(n_r^\prime-k)
  \right]
  \nonumber\\
  &-& E_{nlj} \left( \frac{ Z\alpha }{\eta}-\nu \right)
  (2n_r^\prime-i-k)
  \Biggr\}
  \nonumber\\&\times&
    K_{2,i+k+2\zeta}(\widetilde\kappa_n)
\,,
\end{eqnarray}
where
\begin{eqnarray}
 \nu &=& (-1)^{j+l+1/2} (j+1/2) \,,\nonumber\\
 \zeta &=& \sqrt{\nu^2 -( Z\alpha )^2} \,,\nonumber\\
 \eta &=& \sqrt{1 - (E_{nlj}/m)^2} \,,\nonumber\\
 n_r^\prime &=& n - |\nu| \,,\nonumber\\
 \widetilde\kappa_n&=& n \, \eta \, \kappa_n/( Z\alpha )
 \nonumber
\end{eqnarray} and
\[
E_{nlj}=m\left[1+\frac{( Z\alpha
)^2}{(\zeta+n_r^\prime)^2}\right]^{-1/2}
\]
is the exact relativistic energy of the state for the Dirac-Coulomb
problem. The non-relativistic limit of the relativistic expression
above is given by \eq{ERnljLim} \cite{ejp07}.

\begin{figure}
\begin{center}
\includegraphics[width=0.4\textwidth]{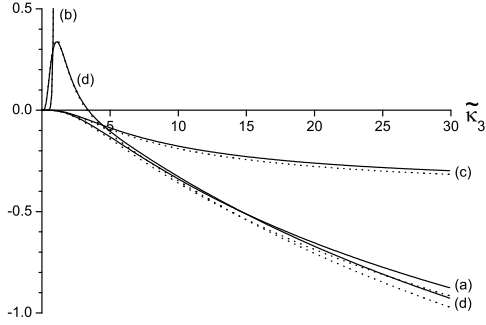}
\end{center}
\caption{Comparison of different approximations of
(\ref{ERnlj_flin}) for $3d_{5/2}$ state: (a) -- the relation
(\ref{ERnlj_flin}), (b) -- the first five terms of expansion for
\eq{Qk}, (c) -- the same for \eq{Qk1k}, (d) -- the same for
\eq{Q11k}. Solid lines correspond to $Z=1$, dotted lines to $Z=92$.
} \label{fig:frelkzi}
\end{figure}

\begin{figure}
\begin{center}
\includegraphics[width=0.3\textwidth]{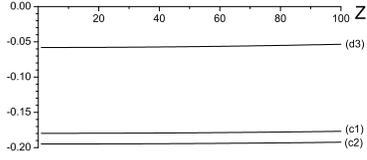}
\end{center}
\caption{Relative errors of different approximations of
(\ref{ERnlj_flin}) for $3d_{5/2}$ state. (c1) -- the first five
terms of expansion for \eq{Qk1k} and $\widetilde\kappa_3=0.3$, (c2)
-- the same for $\widetilde\kappa_3=0.9$, (d3) -- the same for
\eq{Q11k} and $\widetilde\kappa_3=8$. The relative errors of the
other approximations ($a$, $b$, $c$ at $\widetilde\kappa_3=0.3, 0.9,
8$, which are not shown, are above 40\%.} \label{fig:frelzki}
\end{figure}

The efficiency of our approach in the relativistic case is
illustrated for the $3d_{5/2}$ state in Figs.~\ref{fig:frelkzi}
and~\ref{fig:frelzki}.

Our approach for the approximation consists of two parts: firstly, we
approximate a combination of $K_{bc}$ by a presentation with
$_2F_1$. The simplest of such presentations (cf.~\eq{flin}) is
\begin{equation}\label{ERnlj_flin}
 \widetilde K_{bc}(\widetilde\kappa_n) = f(0)\,Q_{bc}(\widetilde\kappa_n) +
 [f(1)-f(0)]\,Q_{b-1,c}(\widetilde\kappa_n)\;.
\end{equation}
Secondly, we approximate the latter by few first terms of the
related $_2F_1$, which may have different arguments in different
presentations and thus is well adjusted for expansion in a
certain region of $\kappa_n$. At both stages we can apply
approximations of the same kind, varying the number of approximation
terms. That improves accuracy, but does not change situation in
general, because we do not change the character of the expressions.

As we have already proved in Sect.~\ref{s:states}, all
non-relativistic results for uncertainty due to the approximation at
the first stage (see, e.g., Table~\ref{tab:ap}) are valid for the
Dirac relativistic consideration.

To study the efficiency at the second step we rely on
(\ref{ERnlj_flin}) as an example. The functions $Q_{bc}(\kappa_n)$ are
typical functions used in the approximations in Table~\ref{tab:ap}
and we approximate $Q_{bc}$, related to $3d_{5/2}$, by first five
terms in (\ref{Qk}), (\ref{Qk1k}) and~(\ref{Q11k}). The results are
summarized in Figs.~\ref{fig:frelkzi} and~\ref{fig:frelzki} and
confirm that the efficiency of the approach does not depend on
whether we proceed relativistically or not. Concerning the value of
the relativistic effects in Figs.~\ref{fig:frelkzi}
and~\ref{fig:frelzki}, we remind that the effects contribute to the
pre-factors in (\ref{ERnlj}), to the index $c=i+k+2\zeta$ of the
$K_{bc}(\widetilde\kappa_n)$ integral and to its argument
$\widetilde\kappa_n$. The figures present only approximations of
$K_{bc}$ at a given argument $\widetilde\kappa_n$, i.e. only a part
of relativistic effects.

\subsection{Corrections to the bound $g$ factor}

As found in \cite{bound_g}, to obtain the correction to the bound
$g$~factor in the case of an arbitrary potential for a Dirac or
Schr\"odinger particle it is enough to know analytically the energy
$E$ at the related level of parametrical accuracy:
\begin{equation}
g_{\rm bound}({nl_j}) = -\frac{\nu}{2j(j+1)} \left[
1-2\nu\,\frac{\partial E_{nlj}}{\partial m} \right]\;.
\end{equation}
Differentiating the function, for which we can numerically control
the accuracy of approximation, we cannot be sure that the numerical
accuracy of the derivative is good enough. However, we expect that the
better is the approximation of the energy the better is the result
for the $g$ factor.

\subsection{Relativistic Uehling correction for a Klein-Gordon particle}

Recently, a perturbative series of the Klein-Gordon bound particle
was discussed \cite{KG1} and the Uehling correction was found for
circular states \cite{cjp05}. It was shown that the Uehling
correction is still expressed in terms of
$K_{bc}(\widetilde\kappa_n^\prime)$, where
\begin{equation}
  \widetilde\kappa_n^\prime
  =
  \kappa_n \left( 1 + \frac{2n-2l-1}{2n^2(2l+1)}(Z\alpha)^2 + \dots \right)\;.
\end{equation}

\subsection{Non-relativistic corrections to the wave function at origin $\Psi(0)$}

An accurate value of the non-relativistic wave function at origin,
$\Psi(0)$, was discussed in muonic and exotic atoms for a number of
occasions. Its value is important for the finite-nuclear-size
corrections in muonic and pionic atoms, for the hyperfine structure
in muonic atoms \cite{ejp98}, for the pionium lifetime
\cite{ejp98,soto} etc.

\begin{figure}
\begin{center}
{\includegraphics[width=0.35\textwidth]{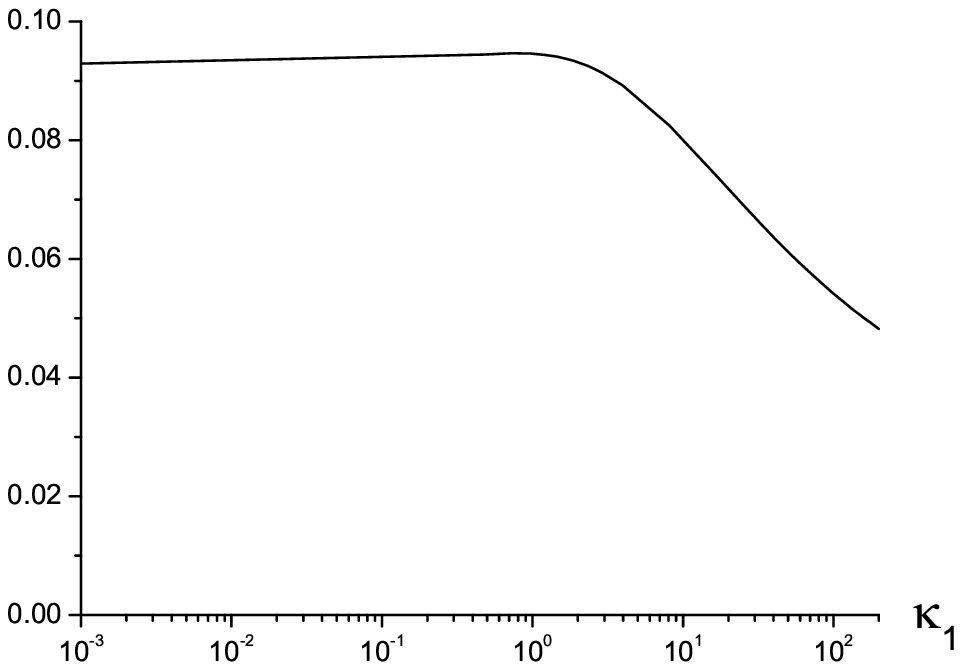}}
{\includegraphics[width=0.35\textwidth]{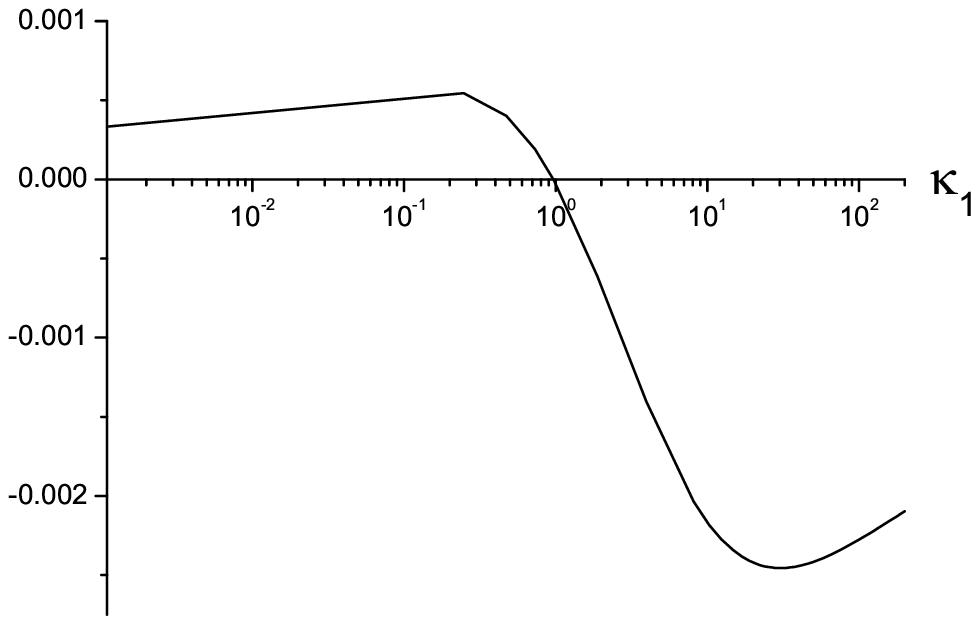}}
\end{center}
\caption{The relative error of ${\delta\Psi_{1s}(0)}/{\Psi_{1s}(0)}$
approximation as a function of $\kappa_1$ for $f=f_0$ (top) and
$f=f_1$ (bottom) from Table~\ref{tab:ap}.}\label{fig:psi0}
\end{figure}

The correction to the wave functions of the $ns$ state
$\Psi_{ns}(0)$ induced by the Uehling potential is of the form
\begin{equation}
  \delta \Psi_{ns}(0) = \int G^\prime_{ns}(0,r) \; V_{\rm U}(r) \; \Psi_{ns}(r) \;
  d^3 r\;,
\end{equation}
where $G^\prime_{ns}(r',r)$ is the non-relativistic reduced Coulomb
Green function. The values of $G^\prime_{ns}(0,r)$ are known in a
simple form (see, e.g., \cite{jetp96}) and in particular
\begin{eqnarray}
  G^\prime_{1s}(0,r)
  &=&
  \frac{m}{2\pi}
  \frac{e^{- Z\alpha  m r}}{r}
  \Big\{2  Z\alpha  m r\bigl[\ln(2  Z\alpha  m r)-\psi(1)\bigr]
  \nonumber\\
  &+& 2( Z\alpha  m r)^2-5 Z\alpha  m r-1\Big\}
  \;.
\end{eqnarray}
For the $1s$ wave function we obtain (cf. \cite{ejp98})
\begin{eqnarray}\label{dpsi0}
  \frac{\delta\Psi_{1s}(0)}{\Psi_{1s}(0)}
  &=&
  \frac{\alpha}{\pi}
  \frac{\kappa_1}{2} \int_0^1 dy\;  f(y) \,\sqrt{1-y}\; (1+\kappa_1 y)^{-3}
  \nonumber\\
  &&\times\;
  \left[ (2+\kappa_1 y) (1+3\kappa_1 y_1) + 2 \kappa_1 y \;\ln \frac{1+\kappa_1 y}{\kappa_1 y}
  \right]
  \nonumber\\
  &=&
  \frac{\alpha}{\pi}
  \Biggl\{
    \frac{1}{\kappa_1^2}
    \biggl[
      K_{43}(\kappa_1)
      +\frac72 \kappa_1 K_{33}(\kappa_1)
      \nonumber\\
      &&+ \frac32 \kappa_1^2 K_{23}(\kappa_1)
    \biggr]
    -
    \left.
    \frac{\partial}{\partial\epsilon}
    K_{2,2+\epsilon}(\kappa_1)
    \right|_{\epsilon=0}
  \Biggr\}
  \,.
\end{eqnarray}
Since the sign of the integrand in \eq{dpsi0} does not
change\footnote{We have found that it also takes place for the $2s$
state.}, we use for its evaluation relations based on
\eq{master_est}, similar to ones obtained above for the energy
corrections. For instance, we can apply the simplest approximations
for $f(y)$ and apply conservative estimation of the uncertainty as
presented in Table~\ref{tab:ap} (see Fig.~\ref{fig:psi0}).

\section{Summary}

Above we have developed a method which allows to study the Uehling
correction to the energy levels and some other atomic
characteristics. The method can be efficiently applied to the
non-relativistic consideration as well as in the relativistic case
for Dirac and Klein-Gordon particles.

The results are presented in terms well-defined at any arguments and
using the well-known hypergeometric function $_2F_1$. The approach
allows to derive various approximations and find useful asymptotics.
The free one-loop vacuum polarization is in principle not a problem
for numerical calculations. Our concern is rather the higher order
corrections and we expect our results can be applied there.

\section*{Acknowledgements}

A part of this work was done during visits of VGI and EYK to
Garching and they are grateful to MPQ for their hospitality. This
work was supported in part by RFBR (grants \#\# 06-02-16156 and
06-02-04018) and DFG (grant GZ 436 RUS 113/769/0-2). The work of EYK
was also supported by the Dynasty foundation.

\appendix

\section*{Appendix}

\section{Transformation of $_2F_1$ adjusted for the $1/\kappa_n$ expansion
\label{as:hkappa}}

The transformation (\ref{Qbc11k}) contains singular terms and in the
text of the paper we consider only the case of $b=2$ (see
(\ref{Q11k})). For integer $b\neq2$ we apply the identity
\begin{eqnarray}
  {}_2F_1(\alpha,\alpha^\prime,\beta,z)
  &=&
  \sum_{k=0}^{N-1}
  \frac{\Gamma(\alpha+k)\Gamma(\alpha^\prime+k)\Gamma(\beta)}{\Gamma(\alpha)\Gamma(\alpha^\prime)\Gamma(\beta+k)}
  \frac{z^k}{k!}
  \nonumber\\
  &+&\frac{\Gamma(\alpha+N)\Gamma(\alpha^\prime+N)\Gamma(\beta)}{\Gamma(\alpha)\Gamma(\alpha^\prime)\Gamma(\beta+N)}
  \frac{z^N}{N!}
  \nonumber\\
  &\times&
  {}_3F_2(\alpha+N,\alpha^\prime+N,1; \; \beta+N, N+1; \; z)
  \nonumber\,.
\end{eqnarray}
to separate the pole of the function for negative integer $\beta$
(we choose for that $N=-\beta+1$). That makes the result for
$Q_{bc}(\kappa_n)$ more complicated, however, with
$z=1/(1+\kappa_n)$ one can easily derive a related expansion for
high $\kappa_n$. The result for integer $b>2$ is
\begin{eqnarray}\label{Qbc11ksum}
  Q_{bc}(\kappa_n)
  &=&
  \left( \frac{\kappa_n}{1+\kappa_n} \right)^c
  \frac{1}{\Gamma(7/2-b)}
  \Biggl\{
  \frac{(-1)^b\sqrt\pi}{2(b-2)!}
  \nonumber\\&&\times\;
  \Biggl[
  {}_2F_1\left(c,3/2,-1+b,\frac{1}{1+\kappa_n} \right)
  \nonumber\\&&\times
  \left[ \ln(1+\kappa_n) -\psi(3/2)-\psi(c)+\psi(b-1)+\psi(1) \right]
  \nonumber\\&+&
  \frac{\partial}{\partial\epsilon}\,
  {}_3F_2\biggl( c-\epsilon,3/2-\epsilon,1;
  \nonumber\\&&\phantom{999999999}
  b-1-\epsilon,1-\epsilon;\;\frac{1}{1+\kappa_n}
  \biggr)\bigg\vert_{\;\epsilon=0}
  \Biggr]
  \nonumber\\&+&
  \frac{(1+\kappa_n)^{b-2}}{\Gamma(c)}
  \sum_{k=0}^{b-3} \Gamma(2-b+c+k) \Gamma\left(\frac72-b+2\right)
  \nonumber\\&&\times
  \frac{(b-k-3)!}{k!}
  \left( -\frac{1}{1+\kappa_n} \right)^k
  \Biggr\}
  \,,
\end{eqnarray}
while for the case of integer $b<2$ we obtain
\begin{eqnarray}\label{Qbc11ksum2}
  Q_{bc}(\kappa_n)
  &=& \left( \frac{\kappa_n}{1+\kappa_n} \right)^c \frac{1}{\Gamma(c)}
  \Biggl\{ \frac{(-1)^b(1+\kappa_n)^{b-2}\Gamma(c+2-b)}{(2-b)!}
  \nonumber\\
  &&
  \times\biggl[{}_2F_1\left( c+2-b, 7/2-b,3-b, \frac{1}{1+\kappa_n} \right)
  \nonumber\\
  &&
  \phantom{9} \times \bigl[\ln(1+\kappa_n)+\psi(1)-\psi(7/2-b)
  \nonumber\\
  &&
  \phantom{999}-\psi(c-b+2)+\psi(3-b) \bigr]
  \nonumber\\
  &&+\frac{\partial}{\partial\epsilon} \; {}_3F_2 \biggl( c+2-b-\epsilon, 7/2-b-\epsilon, 1;\;
  \nonumber\\
  &&\phantom{99999999} 3-b-\epsilon, 1-\epsilon;\; \frac{1}{1+\kappa_n} \biggr) \bigg\vert_{\;\epsilon=0} \biggr]
  \nonumber\\
  &+&\sum_{k=0}^{1-b} \frac{(-1)^k\Gamma(c+k)\Gamma(3/2+k)}{\Gamma(7/2-b)}
  \nonumber\\
  &&\times \frac{(1-b-k)!}{k!} \, \left( \frac{1}{1+\kappa_n} \right)^k
  \Biggr\}\;.
\end{eqnarray}

At $b=2$ the summation in (\ref{Qbc11ksum}) and~(\ref{Qbc11ksum2}) vanishes
($\sum_{k=0}^{b-3}=\sum_{k=0}^{1-b}\equiv 0$) and both
results coincide with \eq{Q11k}.

\section{Estimation of uncertainty of series for $f(y)$ and
related results for the energy shifts \label{as:unc}}

Accordingly to \eq{master_est} and the related discussion, to estimate
the uncertainty of approximation of either a particular base
integral or the correction to the energy for a certain level, it is
enough to find accuracy of the approximation of $f(y)$. Here we
consider its approximation by a partial finite sum of series
(\ref{eithery}) and (\ref{either1y}).

\subsection{Finite $y$-series (\ref{eithery}) and its uncertainty
\label{ass:uncy}}

To estimate uncertainty of the presentation of $f(y)$ by the sum of
the first $N$ terms of \eq{Kbcser0}, we note that the coefficients
of \eq{eithery} satisfy the conditions
\begin{eqnarray}
   \vert c_{k+1}\vert &<& \vert c_k \vert \,,\nonumber\\
  c_k \cdot c_{k+1} &<& 0, \quad k\ge3\;.
\end{eqnarray}
Therefore the fractional uncertainty of the sum is less than
$c_N/f(1)$.  E.g., for the first five terms of (\ref{Kbcser0}) the
relative error for $K_{bc}$ is below 10\%, and for the 14 terms it
is below 1\%.

\subsection{Finite $(1-y)$-series (\ref{either1y}) and its uncertainty
\label{ass:unc1y}}

To estimate uncertainty due to partial summation in~(\ref{Kbcser1}),
we note that
\begin{equation}
  f^\prime_N(0) = \sum_{k=N}^{\infty} c^\prime_k
\end{equation}
and $c^\prime_k<0$ for $k\ge3$. That allows to estimate easily a sum
of coefficients $s^\prime_N=\sum_{k=N}^\infty {c^\prime_k}$ which
enter the remaining series. The difference between the infinite
series for the base integral and the $N$-term partial sum has a
fractional value which does not exceed $s^\prime_N/f(1)$. E.g., the
error of approximation of $K_{bc}$ by the first three terms of
\eq{Kbcser1} is below 20\%, and by the first five terms the error is
below 1\%.

\end{document}